\documentclass[final,3p,times,twocolumn]{elsarticle}

\usepackage{amssymb}
\journal{Results in Physics}

\begin{document}

\begin{frontmatter}

%% Title, authors and addresses

%% use the tnoteref command within \title for footnotes;
%% use the tnotetext command for theassociated footnote;
%% use the fnref command within \author or \address for footnotes;
%% use the fntext command for theassociated footnote;
%% use the corref command within \author for corresponding author footnotes;
%% use the cortext command for theassociated footnote;
%% use the ead command for the email address,
%% and the form \ead[url] for the home page:
%% \title{Title\tnoteref{label1}}
%% \tnotetext[label1]{}
%% \author{Name\corref{cor1}\fnref{label2}}
%% \ead{email address}
%% \ead[url]{home page}
%% \fntext[label2]{}
%% \cortext[cor1]{}
%% \affiliation{organization={},
%%             addressline={},
%%             city={},
%%             postcode={},
%%             state={},
%%             country={}}
%% \fntext[label3]{}

\title{Superfluid transition of the second layer of $^4$He \\ on graphite: does substrate corrugation matter?}

%% use optional labels to link authors explicitly to addresses:
%% \author[label1,label2]{}
%% \affiliation[label1]{organization={},
%%             addressline={},
%%             city={},
%%             postcode={},
%%             state={},
%%             country={}}
%%
%% \affiliation[label2]{organization={},
%%             addressline={},
%%             city={},
%%             postcode={},
%%             state={},
%%             country={}}

\author[inst1]{Massimo Boninsegni}

\affiliation[inst1]{organization={Department of Physics},%Department and Organization
            addressline={University of Alberta}, 
            city={Edmonton},
            postcode={T6G 2E1}, 
            state={Alberta},
            country={Canada}}
            
\author[inst2]{Saverio Moroni}
\affiliation[inst2]{organization={CNR-IOM Democritos and Scuola Internazionale Superiore di Studi Avanzati},%Department and Organization
            addressline={Via Bonomea 265}, 
            city={Trieste},
            postcode={I-34136}, 
            country={Italy}}

\begin{abstract}
The second layer of $^4$He adsorbed on a graphite substrate is studied by Quantum Monte Carlo simulations. We make use of a microscopic model of the substrate fully accounting for its corrugation, and compare the results to those obtained with a smooth substrate. The only effect of corrugation is a $\sim 20$\% reduction of the value of the superfluid fraction of the top layer, in the limit of zero temperature. No evidence of any commensurate (7/12) crystalline and/or ``supersolid'' phase is found; the superfluid transition temperature is estimated to be $\sim 0.75$ K. We discuss the implication of these findings on the interpretation of recent experiments.

\end{abstract}

%%Graphical abstract
%\begin{graphicalabstract}
%\includegraphics{grabs}
%\end{graphicalabstract}

%%Research highlights
%\begin{highlights}
%\item Research highlight 1
%\item Research highlight 2
%\end{highlights}

%\begin{keyword}
%% keywords here, in the form: keyword \sep keyword
%keyword one \sep keyword two
%% PACS codes here, in the form: \PACS code \sep code
%\PACS 0000 \sep 1111
%% MSC codes here, in the form: \MSC code \sep code
%% or \MSC[2008] code \sep code (2000 is the default)
%\MSC 0000 \sep 1111
%\end{keyword}

\end{frontmatter}

%% \linenumbers

%% main text
\section{Introduction}\label{intro}
The possible occurrence of a ``supersolid'' phase in the second adsorbed layer of $^4$He on a graphite substrate continues to elicit much debate. It was first proposed by Greywall and Busch \cite{Greywall1991,Greywall1993}  that the second layer of $^4$He sitting on top of the (solid) first layer, could form a commensurate crystalline phase of two-dimensional (2D) density equal to 4/7 of that of the bottom layer, with a $\sqrt 7 \times \sqrt 7$ partial registry  with respect to it.
Crowell and Reppy \cite{crowell1993,crowell1996} speculated that such a phase might turn superfluid at sufficiently low temperature, giving rise to an intriguing case of a low-dimensional quantum film simultaneously featuring structural and superfluid order\footnote{The denomination {\em supersolid} is strictly speaking not applicable to a system of this type, as the breaking of translational invariance is imposed by an external potential, rather than resulting from the interparticle interactions alone. See, for instance, Ref. \cite{Boninsegni2012}.}.  
\\ \indent
No experimental evidence of a commensurate crystalline phase in the second layer has so far been reported; rather, its occurrence has typically been conjectured based on observed, unexplained anomalies in the temperature behavior of the heat capacity \cite{Greywall1991,Greywall1993,Nakamura2016}; however, such an interpretation of the data has been questioned \cite{Boninsegni2020,Nguyen2021}. Theoretically, the most reliable first principle calculations \cite{Corboz2008,Happacher2013,Moroni2019} have predicted a superfluid phase at low temperature, its superfluid transition conforming to the Berezinskii-Kosterlitz-Thouless (BKT) paradigm (as predicted on other, weaker substrates \cite{Boninsegni1999}), and an incommensurate crystal, the two phases separated by a first-order phase transition. A study has proposed that a commensurate crystal with a slightly different (7/12) density ratio may exist, on top of a first layer of 2D density slightly {\em below} the value at which atomic promotion to second layer is observed \cite{Ahn2016}. \\ \indent
It is unclear what microscopic mechanism should underlie a superfluid transition of a commensurate crystal; a number of theoretical studies have predicted that, while 
in the vicinity of commensuration superfluidity may arise from mobile point defects such as vacancies or interstitials,
the superfluid response should vanish at commensurate density \cite{Boninsegni2001,Prokofev2005,Dang2010}, an effect not observed in experiments
\cite{Nyeki2017,Kim2021}. 
\\ \indent
It is also worth noting that in experiments probing the superfluid response of the second layer, the signal appears at temperatures $T\lesssim 0.3$ K \cite{Kim2021,Nyeki2017b}, whereas computer simulations based on realistic models of the system \cite{Boninsegni2020,Corboz2008,Pierce1998} show a superfluid response whose critical temperature $T_{\rm c}$ is comparable to that of 2D $^4$He (i.e., close to 0.7 K \cite{Boninsegni2006,Boninsegni2006b}), with a superfluid fraction $\rho_S(T)$ approaching 100\% at $T\sim 0.5$ K. Thus, the question may be posed of whether the current microscopic model provides a sufficiently accurate representation of the system of interest, or whether some important physical ingredient is missing (alternatively, of course, one might wonder whether the experimental system is as well-characterized as advertised).
\\ \indent
A specific point of contention, not yet fully explored but which might conceivably account for the discrepancy between theory and experiment, is the role of the corrugation of the graphite substrate. While it is accepted that its explicit inclusion in the microscopic model of the system is necessary, in order to capture the existence of a commensurate crystalline phase in the {\em first} adsorbed layer \cite{Corboz2008,Pierce2000}, it has been generally assumed that its effect on the second and higher $^4$He adlayers is negligible, and therefore an acceptable microscopic description may be achieved assuming the graphite substrate to be featureless (i.e., flat) \cite{Hernandez2003,Turnbull2005}. 
\\ \indent
The  main arguments in support of this choice are the relatively large average distance from the substrate of the helium atoms in the second layer, as well as the expectation that the underlying crystalline $^4$He layer, which is incommensurate with the graphite substrate, should largely screen its effect on helium atoms in other layers. On the other hand, the formation of a commensurate crystalline phase at low temperature may hinge on a delicate energy balance, possibly sensitive to the finer details of the substrate structure \cite{Ahn2016}; furthermore, the experimentally observed suppression of $T_{\rm c}$, with respect to what is predicted theoretically, is consistent with a significant, externally imposed modulation of the film \cite{Leggett1970}.
\\ \indent
In order to clarify this outstanding issue, we have carried out extensive Quantum Monte Carlo (QMC) simulations of a 2-layer film of $^4$He adsorbed on graphite, making use of an accepted 
{\em ab initio} potential describing the interaction of a $^4$He atom with the substrate, specifically designed to capture the effects of substrate corrugation. This potential has been adopted in all previous simulations studies. Specifically, we study the superfluid transition of the second layer, for the particular case in which its 2D density is 7/12 of the first layer. 
\\ \indent
Our results show once again {\em no evidence} of a commensurate crystalline phase in the second layer, nor of any ``supersolid'' phase. Rather, the system undergoes a BKT superfluid transition, with a critical temperature $T_{\rm c}=0.74(2)$ K, i.e., still much higher than what observed experimentally, despite the explicit inclusion of substrate corrugation in the microscopic Hamiltonian. Comparison of the results with those yielded by a flat graphite substrate shows that the only effect of the corrugation is a relatively small suppression ($\sim$20\% at $T=0$) of the superfluid fraction of the top layer, with no significant enhancement of structural order. \\ 
\indent
Thus, barring some unexpected scenario in which the helium-graphite potential or other parts of the microscopic model utilized in this work (which is standard) should turn out to be seriously deficient, it appears as if the contention of a possible supersolid phase and/or the interpretation of the existing experimental data may require reconsideration.
The remainder of this paper is organized as follows: in Sec. \ref{mod} we describe the microscopic model of the system adopted in this work, which is the same with that used in previous studies \cite{Corboz2008,Pierce1998,Pierce2000}; in Sec. \ref{meth} we briefly describe our methodology; we illustrate our results in Sec. \ref{res}, and outline our conclusions in Sec. \ref{conc}.

\section{Model}\label{mod}
We consider an ensemble of $N$ $^4$He atoms, regarded as point-like spin-zero bosons, moving in the presence of a graphite substrate, modeled as described below. 
\begin{figure}[h]
\centering
\includegraphics[width=\linewidth]{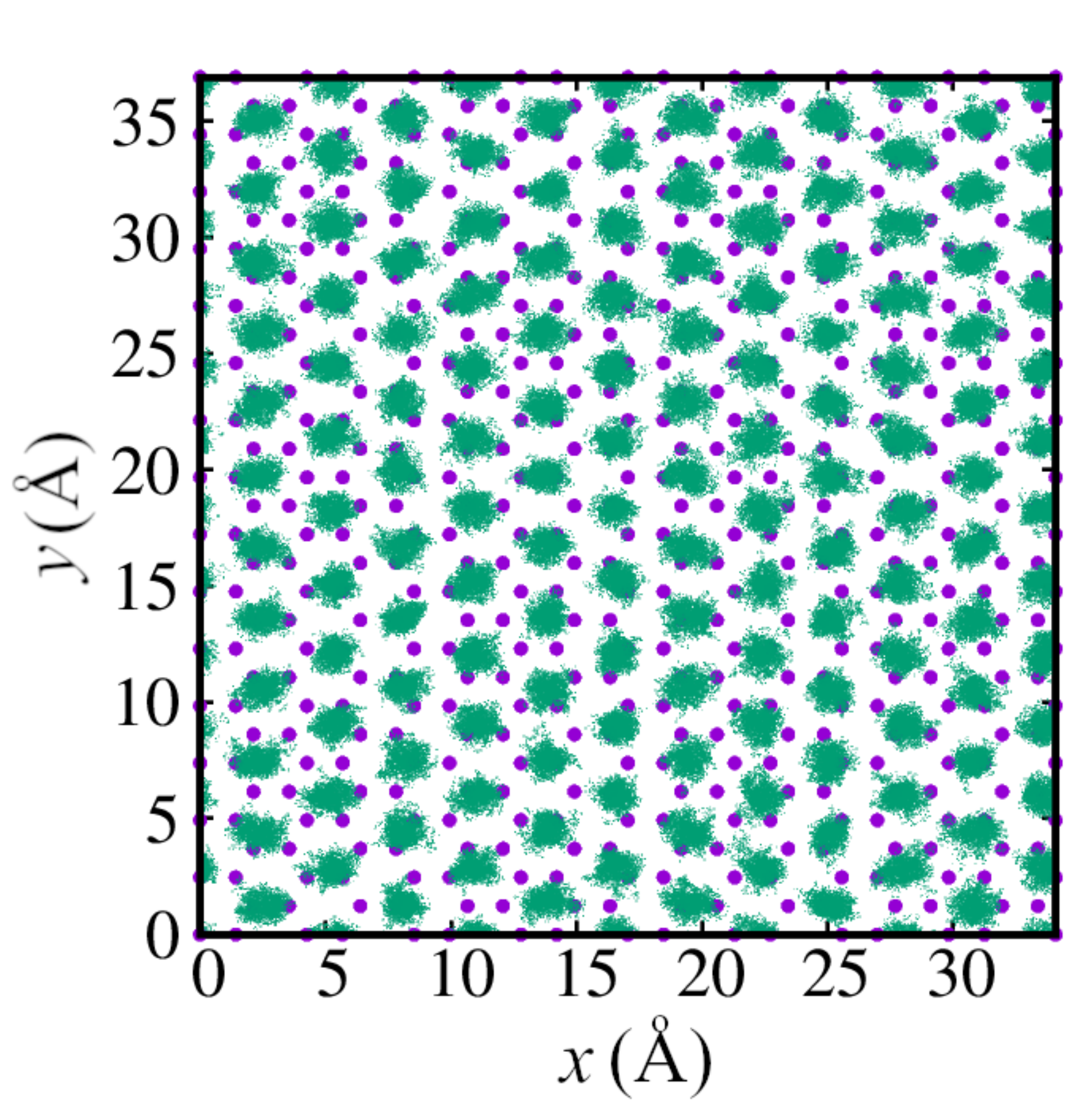}
\caption{Simulation cell in the $x$-$y$ plane. Carbon atoms are shown by (480) solid circles, while fuzzy ``clouds'' are the traces (many-particle world lines)  of  the (144) $^4$He atoms in the first adlayer. This particular snapshot pertains to a simulation at temperature $T=0.4$ K.}
\label{f1}
\end{figure}
The system is enclosed in a simulation cell of sizes $L_x\times L_y\times L_z$,  with periodic boundary conditions in all directions (but $L_z$ is taken large enough to make boundary conditions in the $z$ direction irrelevant). The graphite substrate lies on the $z=0$ plane, and we choose $L_x=34.08$ \AA, $L_y=36.8927$ \AA, which allows us to fit exactly 480 C atoms in the simulation cell (see Fig. \ref{f1}), without introducing any unwanted frustration at the boundaries (which could quite conceivably result in a spurious superfluid signal).
\\ \indent
The nominal coverage $\theta$ is given by $N/A$.
In this work, the number of particles $N$ is set to 228; of these, $144$ occupy the first layer, forming a triangular, incommensurate solid film\footnote{The lattice arrangement of $^4$He atoms shown in Fig. \ref{f1} arises {\em spontaneously} in a simulation with just 144 atoms.} (shown in Fig. \ref{f1}) of coverage $\theta_L=0.1145$ \AA$^{-2}$, while the remaining 84 constitute the second layer of 2D density equal to $\theta_U= 7\theta_L/12=0.0668$ \AA$^{-2}$, so that the total coverage $\theta=\theta_L+\theta_U=0.181$ \AA$^{-2}$. 
\\ \indent
The coverage $\theta_L$ is very close to that at which atomic promotion to the second layer is observed in most simulation work \cite{Corboz2008,Ahn2016,Whitlock1998}, as well as in experiments \cite{Bretz1973,Polanco1978,Carneiro1981,Lauter1987}. Upon adsorbing a second layer on top of it, the first layer is  compressed, but the effect is relatively small (it can be estimated at $\sim$ 1\% at the value of $\theta_U$ considered in this work \cite{Lauter1987}), and therefore we neglect it here, for the purpose of keeping the simulation cell commensurate with the graphite substrate, without introducing vacancies or interstitials in the first $^4$He adlayer.
\\ \indent
The quantum-mechanical many-body Hamiltonian reads as follows:
\begin{eqnarray}\label{u}
\hat H = -\sum_{i}\lambda\nabla^2_{i}+\sum_{i<j}v(r_{ij})+\sum_{i}V({\bf r}_{i}).
\end{eqnarray}
The first and third sums run over all the $N$ $^4$He atoms,  $\lambda=6.0596$ K\AA$^{2}$; the second sum runs over all pairs of particles, $r_{ij}\equiv |{\bf r}_i-{\bf r}_j|$, 
${\bf r}_i\equiv(x_i,y_i,z_i)$ being the position of the $i$th atom, and $v(r)$ is the accepted Aziz pair potential \cite{Aziz1979}, which describes the interaction between two helium atoms. 
$V$ is the potential describing the interaction of a helium atom with the graphite substrate; for consistency with virtually {\em all} previous theoretical studies, we use the 
%SM
anisotropic 6--12
Carlos-Cole potential \cite{Carlos1980}, which can be expressed as follows:
\begin{equation}\label{carloscole}
    V({\bf r}) = V_0(z) + \sum_{\bf G} V_{\bf G}(z)\ {\rm exp}\  [ i\ (G_xx+G_yy)]\ .
\end{equation}
$V_0(z)$ is a term that only depends on the distance of the atom from the basal plane; it corresponds to the laterally averaged potential normally utilized to represent a flat substrate \cite{Boninsegni2020,Pierce2000,Whitlock1998}.  The sum in the second term of the right-hand side of (\ref{carloscole}) runs over all graphite reciprocal lattice vectors {\bf G}$\equiv (G_x,G_y,0)$ of the graphite substrate. The functions $V_0(z), V_{\bf G}(z)$ are provided in Ref. \cite{Carlos1980}. We come back to the details of the evaluation of (\ref{carloscole}) in our simulations in Sec. \ref{meth}.
\\ \indent
At the $^4$He coverage and in the temperature range considered here, two distinct atomic layers form.  
In principle, of course, $^4$He atoms are identical, and therefore no conceptual distinction can be drawn between atoms in the ``top'' and ``bottom'' layer. However, it has been consistently found in previous work (see, for instance, Ref. \cite{Boninsegni2020}) that both inter-layer hopping of atoms, as well as quantum-mechanical exchanges among atoms in the first adsorbed  layer (which orders as a triangular crystal) and/or in different layers, are {\em exceedingly infrequent}. It is therefore an  excellent approximation to regard atoms in the bottom layer as {\em distinguishable} quantum particles (i.e., ``Boltzmannons''); on the other hand, atoms in the top layer are considered as indistinguishable, and can therefore undergo quantum exchanges. 
%\begin{figure}[h]
%\centering
%\includegraphics[width=\linewidth]{scheme.pdf}
%\caption{Schematic of the simulation setup. The simulation  cell is elongated, sandwiched between %homogeneous solid and superfluid phases,
%modeled as homogeneous, continuous media.}
%\label{scheme}
%\end{figure}

\section{Methodology}\label {meth}
The QMC methodology adopted here is the canonical \cite{Mezzacapo2006,Mezzacapo2007} version of the continuous-space Worm Algorithm \cite{Boninsegni2006,Boninsegni2006b}, a finite temperature ($T$) quantum Monte Carlo (QMC) technique. 
Details of the simulations carried out in this work are standard, and therefore the reader is referred to the original references. All of the results quoted here are extrapolated to the limit of time step $\tau\to 0$.
\\ \indent
The calculation of the helium-graphite potential (Eq. \ref{carloscole}) in the simulation has been performed in two different ways, which, as we have verified, yield the same results within statistical errors. In the first scheme, we evaluated (\ref{carloscole}) by tabulating and interpolating the functions $V_0(z)$ and $V_{\bf G}(z)$, and by performing a direct, on-the-fly summation over terms associated to a subset of reciprocal lattice vectors; we found that the sum converges rather rapidly, and that he inclusion of the twelve shortest reciprocal lattice vectors is sufficient to achieve a numerically accurate representation of $V({\bf r})$. In the second scheme
%SM
the cutoff in reciprocal space is avoided by tabulating and interpolating
in $x,y,z$ the real-space expression \cite{Carlos1980} of the He-C pair potential summed on individual C atoms. The number of substrate atoms
included in the tabulation ensures convergence of the helium-graphite 
potential to better than 0.01 K.
%MB Saverio, potresti scrivere tu qui cosa fai?

The key physical quantity computed in this work is the superfluid fraction $\rho_S(T)$ of the top layer as a function of temperature, for which we use the well-known winding number estimator \cite{Ceperley1995}. We also evaluate energetic and structural properties, such as  density profiles and the pair-correlation function $g(r)$, integrated along the direction $z$,  perpendicular to the substrate, for both layers separately.  Crystalline order in both layers may also be monitored through the visual inspection of the imaginary-time paths. 
\section{Results}\label{res}
\begin{figure}[h]
\centering
\includegraphics[width=\linewidth]{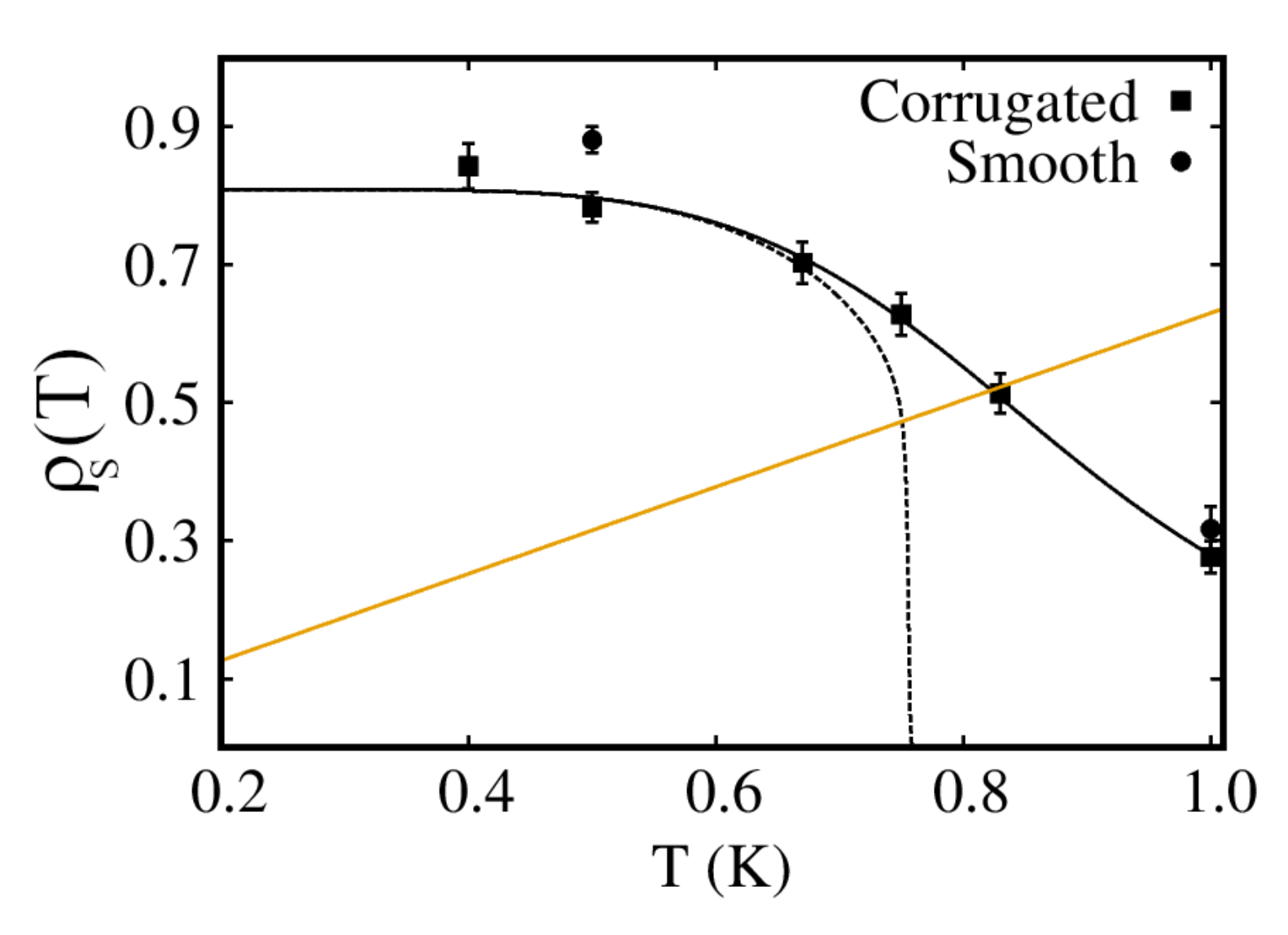}
\caption{Superfluid fraction of the top $^4$He layer as a function of temperature. Boxes are estimates obtained on a corrugated substrate, while circles are for a smooth substrate. Solid line is a fit to the data based on the BKT recursive equations (see text). Dotted line represents the extrapolation to the thermodynamic limit, while the straight line through the origin is the universal jump condition.}
\label{f2}
\end{figure}
The main result of this work is shown in Fig. \ref{f2}, displaying the estimates for the superfluid fraction $\rho_S(T)$ as a function of temperature, in the temperature interval $0.4 \le T\le 1$ K. We compare the values obtained on the fully corrugated substrate (boxes) with those on a smooth, flat substrate, which corresponds to setting $V({\bf r})=V_0(z)$ (Eq. \ref{carloscole}). \\ \indent
The fit to the data on a corrugated substrate is obtained following the procedure outlined in Ref. \cite{Pollock1987}, with the only difference that we have to allow the value of $\rho_S(T=0)$ to be treated as a free parameter, in order to fit our data. In other words, the corrugation of the graphite substrate induces a spatial modulation of the superfluid (i.e., the breaking of translational invariance caused by an external agent), causing the superfluid fraction to saturate to less than 100\% in the $T\to 0$ limit, an effect predicted a long time ago \cite{Leggett1970} and already observed in other systems \cite{Boninsegni2012b}. On comparing the results obtained in the presence and in the absence of corrugation (the latter shown by filled circles in Fig. \ref{f2}), one can see that this relatively small suppression of the superfluid signal in the low temperature limit is the sole quantitative effect of corrugation.
\\ \indent
The optimal value of the fitting parameter $\rho_S(T=0)$ is 0.82(2). Based on the fit to the data for our finite system, we can infer the behavior in the thermodynamic limit, shown by the dotted curve in Fig. \ref{f2}, and with the aid of the well-known universal jump condition \cite{Nelson1977}, we estimate the superfluid transition temperature $T_c=0.74(2)$ K, i.e., close to that of other $^4$He monolayer systems \cite{Boninsegni1999}. 
\begin{figure}[h]
\centering
\includegraphics[width=\linewidth]{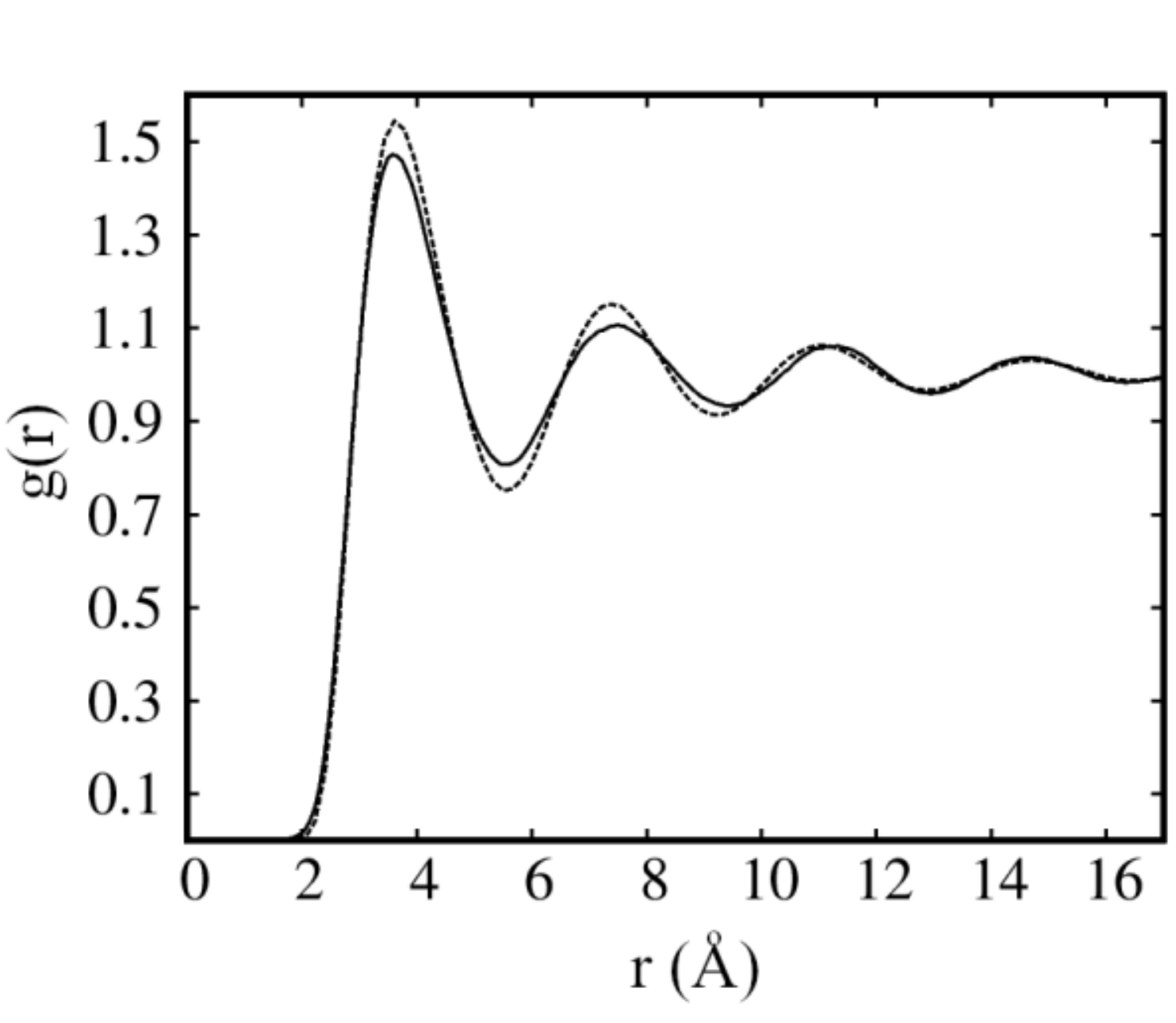}
\caption{Pair correlation function $g(r)$ for the upper layer of a $^4$He film of coverage $\theta_U=0.0668$ \AA$^{-2}$, integrated over the axis perpendicular to the substrate. These results shown pertain to a temperatures $T$ =0.4 K. Also shown (dotted line) is the result for a purely 2D $^4$He film at the same 2D density and temperature.}
\label{f3}
\end{figure}
\\ \indent
A quantitative assessment of the effect of the inclusion of corrugation of the graphite substrate on the physical properties of the 2-layer $^4$He film is provided by a comparison not only of the superfluid response, but also of energetics, computed using the same methodology but with the two different models of the substrate (smooth and corrugated). For example, the kinetic energy per $^4$He atom in the {\em first} layer at $T=0.5$ K, computed using the full model (\ref{carloscole}),
is equal to 55.43(6) K, i.e.,  very close to the value of 55.18(9) K on a smooth substrate, i.e., using the laterally averaged $V_0(z)$; for the $^4$He atoms in the top layer, the result is the same for the two cases,  within statistical errors, namely 17.1(1) K. Together with the results for the superfluid fraction, all of this provides quantitative confirmation for the long held belief that substrate corrugation does not substantially affect the physics of a superfluid $^4$He monolayer.
\\ \indent
Structurally, no evidence is seen of the formation of a commensurate crystalline layer at low temperature. Fig. \ref{f3} displays the reduced pair correlation function for  the second layer of $^4$He, at the lowest temperature considered here, namely $T=0.4$ K (no discernible temperature dependence is observed in this study). The pair correlation function shows  the rapidly decaying oscillations that are typical of a fluid. \\ \indent 
It is interesting to compare this function to the pair correlation of purely 2D $^4$He at the same nominal density, namely $\theta_U$, shown by dotted lines in Fig. \ref{f3} (this function was computed in this work in a separate simulation). The 2D system features noticeably more pronounced oscillations, as in 2D this particular value of the density is relatively close to freezing \cite{Gordillo1998}. On the other hand, motion in the direction perpendicular to the substrate acts to soften the repulsive part of the interatomic potential, in the top layer of an adsorbed $^4$He film, therefore strengthening the fluid phase. Nonetheless, the superfluid fraction of the 2D system for which results are shown in Fig. \ref{f3} is 100\%, within statistical errors, which is consistent with the absence of any external physical mechanism that breaks translational invariance.

\section{Discussion and Conclusions}\label{conc}
The effect of surface corrugation on the physics of the second layer of $^4$He adsorbed on graphite was investigated by first principle numerical simulations. We made use of the accepted microscopic pair potential describing the interaction of a helium atom with a graphite substrate, ubiquitously adopted in the past to study, for example, the phase diagram of the adsorbed first layer of $^4$He \cite{Pierce2000}. 
\\ \indent
The results of this study are inconsistent with recent experimental measurements, in which a substantially lower superfluid transition temperature is observed (roughly a factor two lower than that reported here).
It is certainly possible that the microscopic model utilized here may be seriously inadequate, and that the effect of the corrugation may be much stronger than assessed here. This seems highly unlikely though; while the quantitative accuracy of the Carlos-Cole interaction has not yet been definitively established \cite{Badman2018}, nonetheless it successfully accounts for many of the experimentally observed features of the first layer. 
It is also important to note that virtually all experimental works in which the claim is made of a possible ``supersolid'' phase in the second layer, assume the correctness of an old phase diagram computed using a different QMC methodology but the same potential used here. While it is certainly important to continue to refine and improve our microscopic model of the helium-graphite interaction, it also seems equally appropriate to investigate the scenario in which the experimental substrate may be considerably more imperfect than perhaps assumed so far, which would undoubtedly be consistent with the significant reduction of the transition temperature, with respect to what theoretically estimated.
\\ \indent
Our results are not qualitatively different from those obtained assuming a flat substrate. No evidence is seen of the formation of a commensurate crystal at low temperature, at least down to $T=0.4$ K. Obviously, one could always contend that the crystal (of whose existence, it is worth restating, no experimental evidence whatsoever has yet been produced) may form at temperatures lower than those considered here. The scenario of a supersolid melting into a superfluid at constant density cannot be excluded \cite{Boninsegni2005}, but no evidence of it has been observed when crystallization takes place due to an external ``pinning'' potential \cite{Dang2010}. In general, however, without a clear physical mechanism and/or quantitative energy balance suggesting that crystallization might indeed take place at some reasonably well-defined low temperature, this hypothesis seems not falsifiable, hence unscientific.
\\ \indent
In any case, the results of this study reaffirm the conventional knowledge, namely that the external imposition of a density modulation on a superfluid does not bring about any new or unexpected physical behavior, a small suppression of the superfluid response being the only significant signature.

\section{Declaration of Competing Interests}
The authors declare no known competing financial interests or personal relationships that may appear to influence the work reported in this paper.

\section{Acknowledgments}
This work was supported by the Natural Sciences and Engineering Research Council of Canada.

 \bibliographystyle{elsarticle-num} 
 \bibliography{refs}

%% else use the following coding to input the bibitems directly in the
%% TeX file.

% \begin{thebibliography}{00}

% %% \bibitem{label}
% %% Text of bibliographic item

% \bibitem{}

% \end{thebibliography}
\end{document}